\documentclass[aps,twocolumn]{revtex4}

\usepackage{amssymb}
\usepackage{amsmath}
\usepackage{bm}
\usepackage[dvips]{epsfig}
\usepackage{graphicx}

\begin{document}

\title{Poisson brackets of hydrodynamic type and their generalizations.}

\author{A.Ya. Maltsev$^{1}$, S.P. Novikov$^{1,2}$}

\affiliation{
\centerline{$^{1}$ \it{L.D. Landau Institute for Theoretical Physics}}
\centerline{\it 142432 Chernogolovka, pr. Ak. Semenova 1A}
\centerline{$^{2}$ \it{V.A. Steklov Mathematical Institute 
of Russian Academy of Sciences}}
\centerline{\it 119991 Moscow, Gubkina str. 8}
}

\begin{abstract}
 In this paper, we consider Hamiltonian structures of hydrodynamic type 
and some of their generalizations. In particular, we discuss the questions 
concerning the structure and special forms of the corresponding Poisson 
brackets and the connection of such structures with the theory of 
integration of systems of hydrodynamic type.
\end{abstract}

\maketitle

\vspace{5mm}

\section{Introduction}

 This work will be devoted to Hamiltonian structures, which currently 
play an important role in many areas of mathematics and mathematical 
physics. Namely, we will mainly consider here Hamiltonian structures 
of hydrodynamic type and some of their important generalizations.
Traditionally, the Hamiltonian structures of the hydrodynamic type 
are associated with the Poisson brackets arising in hydrodynamics 
and representing the brackets for the corresponding hydrodynamic 
densities. Brackets of this type represent, as a rule, expressions 
of the first order in spatial derivatives and can be written in the 
following general form
\begin{equation}
\label{GenBr}
\{ U^{\nu} ({\bf x}) , U^{\mu} ({\bf y}) \} \,\,\, =   
\end{equation}
$$ = \,\,\, g^{\nu\mu i} \left( {\bf U} ({\bf x}) \right) \,\,
\delta_{x^{i}} ({\bf x} - {\bf y}) \,\, + \,\, 
b^{\nu\mu i}_{\lambda} \left( {\bf U} ({\bf x}) \right) 
U^{\lambda}_{x^{i}} \,\, \delta ({\bf x} - {\bf y}) \,\,\, , $$
where $\, {\bf U} ({\bf x}) \, $ is a complete set of hydrodynamic
densities in the problem under consideration.

 Brackets (\ref{GenBr}) in hydrodynamics are usually associated with 
Hamiltonians of hydrodynamic type, i.e. Hamiltonians of the form
\begin{equation}
\label{GenHam}
H \,\,\, = \,\,\, \int P_{H} \left( {\bf U} ({\bf x}) \right)
\,\, d^{n} x \,\,\, , 
\end{equation}
where $\, n \, $ is the dimension of the problem under consideration.
It is easy to see that the Hamiltonians ({\ref{GenHam}) in the 
Hamiltonian structure (\ref{GenBr}) correspond to systems of the form
\begin{equation}
\label{GenSyst}
U^{\nu}_{t} \,\,\, = \,\,\, V^{\nu i}_{\mu} 
\left( {\bf U} ({\bf x}) \right) \, U^{\mu}_{x^{i}} \,\,\, ,
\end{equation}
$$\nu , \, \mu \,\, = \,\, 1 , \dots , N \, , \quad
i \,\, = \,\, 1 , \dots , n $$

 The bracket (\ref{GenBr}) and the system (\ref{GenSyst}) are written 
in the most general form, which does not reflect any specifics of 
hydrodynamic variables $\, {\bf U} ({\bf x}) \, $. 
It is easy to see that the presented form is invariant with respect 
to any ``point'' change of variables
$\, {\tilde {\bf U}} \, = \, {\tilde {\bf U}} ({\bf U}) \, $
with appropriate transformations of the quantities
$\, g^{\nu\mu i} ({\bf U}) $, $\, b^{\nu\mu i}_{\lambda} ({\bf U}) \, $
and $\, V^{\nu i}_{\mu} ({\bf U}) \, $. 
At the same time, certainly, in real hydrodynamics, each variable, 
as a rule, has its own special physical meaning, and the corresponding 
brackets (\ref{GenBr}) and systems (\ref{GenSyst}) have an additional 
structure associated with this fact. As is well known (\cite{Landau}), 
in the simplest case of a barotropic flow of an ideal fluid, the Poisson 
brackets of the fluid density $\, \rho ({\bf x}) \, $ and its velocity 
components $\, v^{i} ({\bf x}) \, $ can be written as
\begin{multline}
\label{HydrSkobka}
\{ \rho ({\bf x}) , \rho ({\bf y}) \} \,\,\, = \,\,\, 0 \,\, ,
\quad \{ v^{i} ({\bf x}) , \rho ({\bf y}) \} \,\,\, = \,\,\,
\nabla_{x^{i}} \, \delta ( {\bf x} - {\bf y} )  \,\, ,  \\
\{ v^{i} ({\bf x}) , v^{k} ({\bf y}) \} \,\,\, = \,\,\,
{1 \over \rho ({\bf x})} \, \left( 
{\partial v^{k} \over \partial x_{i}} \, - \, 
{\partial v^{i} \over \partial x^{k}}  \right) \,
\delta ( {\bf x} - {\bf y} ) \,\,\, ,
\end{multline}
and the Hamiltonian corresponding to such a flow has the form
$$H \,\,\, = \,\,\, \int \left( {\rho v^{2} \over 2} \, + \,
\epsilon (\rho) \right) \, d^{3} x $$

 An extremely important property of the bracket (\ref{HydrSkobka}) 
is that the condition
\begin{equation}
\label{ZeroRot}
{\rm rot} \, {\bf v} ({\bf x}) \,\,\ = \,\,\, 0 
\end{equation}
is conserved for any Hamiltonian fluid dynamics.
Determining in this case the flow potential $\, \Phi ({\bf x}) \, $
according to the standard formula
$$\, {\bf v} ({\bf x}) \,\,\ = \,\,\, 
\nabla \, \Phi ({\bf x}) \,\,\, , $$
it is easy to check that the variables
$\, \rho ({\bf x}) \, $ and $\, \Phi ({\bf x}) \, $
define the canonical Poisson bracket
$$\{ \rho ({\bf x}) , \rho ({\bf y}) \} \,\,\, = \,\,\, 0 \,\, ,
\quad \{ \Phi ({\bf x}) , \Phi ({\bf y}) \} \,\,\, = \,\,\, 0 
\,\, ,  $$
\begin{equation}
\label{StandardSkobka}
\{ \Phi ({\bf x}) , \rho ({\bf y}) \} \,\,\, = \,\,\, 
 \delta ( {\bf x} - {\bf y} )
\end{equation}

 It can also be noted that, in their physical meaning, the variables
$\, \rho ({\bf x}) \, $ can be classified as action-type variables, 
and the variables $\, \Phi ({\bf x}) \, $ as angular (phase) variables.
It is also well known that the introduction of canonical variables 
for the bracket (\ref{HydrSkobka}) in the more general case of vortex 
flows is more complicated and is related to the definition of Clebsch 
variables for such flows (see \cite{Lamb}).

 Hamiltonian structures (\ref{GenBr}) also arise for more general case 
of non-barotropic flows, as well as for the equations of 
magnetohydrodynamics (\cite{MorrisonGreene}). It can be shown that, 
both in the barotropic and in the more general non-barotropic case, 
the introduction of canonical variables (Clebsch variables) for 
Hamiltonian structures is related to the representation of the 
corresponding equations in the form of a constrained Lagrangian 
system (see \cite{Davydov,Khalatnikov}). Moreover, this approach 
turns out to be fruitful also in the description of non-entropic flows 
of a classical fluid, as well as superfluidity
(\cite{PokrovskyKhalatnikov}). It can also be noted that in the 
latter case, the Lagrangian and Hamiltonian approaches often turn out 
to be an important component not only in describing certain aspects of 
the dynamics of a superfluid fluid, but also in establishing the 
equations of such dynamics in general (see, for example.
\cite{LebedevKhalatnikov1,KhalatnikovLebedev,LebedevKhalatnikov2,
VolovikDotsenko1,VolovikDotsenko2}). In the general case, as is well 
known, the construction of canonical variables for Hamiltonian 
structures in hydrodynamics is a very important problem associated 
with the description of many features of the corresponding flows, 
including their topological features
(see \cite{KuznetsovMikhailov,ZakharovKuznetsov}).

 Below we will show in fact that in many cases for brackets of 
hydrodynamic type, it seems natural to have a more extended 
definition of the canonical form. Moreover, in addition to the 
canonical form of the bracket (\ref{GenBr}), another (diagonal) 
form of this bracket also turns out to be extremely important in 
the study of the corresponding Hamiltonian systems. The latter 
circumstance will be most obvious in the case of one spatial dimension, 
where the theory of such brackets (and their generalizations) 
represents the basis of the theory of integrable systems of 
hydrodynamic type. As is also well known, other structures of 
Poisson brackets (\ref{GenBr}) are often also important, 
in particular, their Lie algebraic structure (see eg.
\cite{Arnold,ArnoldKozlovNeishtadt,DubrovinNovikov}).

 Generalizations of Hamiltonian structures of hydrodynamic type 
include structures containing simultaneously hydrodynamic and 
phase variables, structures combining hydrodynamic and 
Lie-algebraic parts, structures containing higher derivatives 
and nonlocal additives, etc. A huge variety of extremely important 
structures of this type were considered by I.E. Dzyaloshinsky 
and G.E. Volovik in the work \cite{DzyaloshinskiiVolovik}. 
In particular, as was shown in \cite{DzyaloshinskiiVolovik}, 
a huge variety of applications of Hamiltonian structures 
generalizing structures of hydrodynamic type include the 
description of elastic dynamics of crystals with impurities 
and defects, the description of the dynamics of liquid crystals, 
the dynamics of magnets of various types, as well as spin glasses, 
etc. This work is dedicated to the 90th anniversary of 
I.E. Dzyaloshinsky.

\section{One-dimensional Hamiltonian structures of hydrodynamic type}
\setcounter{equation}{0}

 In the case of one spatial variable, systems of hydrodynamic 
type have the form
\begin{equation}
\label{OneDimSyst}
U^{\nu}_{t} \,\,\, = \,\,\, V^{\nu}_{\mu} 
\left( {\bf U} \right) \, U^{\mu}_{x} 
\end{equation}

 The matrix $\, V^{\nu}_{\mu} ({\bf U}) \, $ represents 
a linear transformation matrix on the tangent space of a 
manifold with coordinates $\, {\bf U} \, $, in particular, 
it has the appropriate transformation law under the point
transformations
$\, {\tilde {\bf U}} \, = \, {\tilde {\bf U}} ({\bf U}) \, $.

 A system (\ref{OneDimSyst}) is hyperbolic in some range 
of values of $\, {\bf U} \, $ if at each point of this domain 
all eigenvalues of $\, V^{\nu}_{\mu} ({\bf U}) \, $ are real, 
and the corresponding eigenvectors form a basis in the tangent space.
A system (\ref{OneDimSyst}) is called strictly hyperbolic in some 
domain if at each point of this domain the eigenvalues of
$\, V^{\nu}_{\mu} ({\bf U}) \, $ are real and pairwise are different.

 In the case of two-component systems (that is, 
$\, {\bf U} = (U^{1}, U^{2})$), each strictly hyperbolic system 
(\ref{OneDimSyst}) can be reduced to diagonal form
\begin{equation}
\label{DiagSyst}
R^{\nu}_{t} \,\,\, = \,\,\, v^{\nu} 
\left( {\bf R} \right) \, R^{\nu}_{x} 
\end{equation}
with a real change of variables
$\, {\bf R} \, = \, {\bf R} ({\bf U}) \, $.
In the case of $\, N \geq 3 \, $, however, such a reduction, 
generally speaking, is not always possible. In general, 
a strictly hyperbolic system (\ref{OneDimSyst}) can be locally 
reduced to a diagonal form (using a real change of coordinates) 
if the corresponding Hantjes tensor is identically zero.
In coordinate form, the components of the Hantjes tensor can be 
written as
\begin{multline*}
H^{\nu}_{\mu\lambda} ({\bf U}) \,\,\, =  \\
= \,\,\, V^{\nu}_{\sigma} ({\bf U}) \, V^{\sigma}_{\tau} ({\bf U}) \,
N^{\tau}_{\mu\lambda} ({\bf U}) \, - \,
V^{\nu}_{\sigma} ({\bf U}) \, N^{\sigma}_{\tau\lambda} ({\bf U}) 
\, V^{\tau}_{\mu} ({\bf U}) \, -  \\
- \, V^{\nu}_{\sigma} ({\bf U}) \, N^{\sigma}_{\mu\tau} ({\bf U}) 
\, V^{\tau}_{\lambda} ({\bf U}) \, + \, 
N^{\nu}_{\sigma\tau} ({\bf U})  \, V^{\sigma}_{\mu} ({\bf U}) \,
V^{\tau}_{\lambda} ({\bf U}) \,\,\, ,
\end{multline*}
where $\, N^{\nu}_{\mu\lambda} ({\bf U}) \, $ is the Nijenhuis 
tensor of the operator $\, V^{\nu}_{\mu} ({\bf U}) \, $:
\begin{multline*}
N^{\nu}_{\mu\lambda} ({\bf U}) \,\,\, = \,\,\, 
V^{\sigma}_{\mu} ({\bf U}) \,
{\partial V^{\nu}_{\lambda} \over \partial U^{\sigma}} \,\,\, - 
\,\,\, V^{\sigma}_{\lambda} ({\bf U}) \,
{\partial V^{\nu}_{\mu} \over \partial U^{\sigma}} \,\,\, +  \\
+ \,\,\, V^{\nu}_{\sigma} ({\bf U}) \,
{\partial V^{\sigma}_{\mu} \over \partial U^{\lambda}} \,\,\, -
\,\,\, V^{\nu}_{\sigma} ({\bf U}) \,
{\partial V^{\sigma}_{\lambda} \over \partial U^{\mu}}
\end{multline*}

 In the most general case, functions  $\, R ({\bf U}) \, $,
satisfying, by virtue of the system (\ref{OneDimSyst}), 
the equation 
$$R_{t} \,\,\, = \,\,\, v ({\bf U}) \, R_{x} $$
for some function $\, v ({\bf U}) \, $, are called the Riemann 
invariants of the system (\ref{OneDimSyst}). In the general case, 
system (\ref{OneDimSyst}) may have a certain set of independent 
Riemann invariants, the number of which, however, is insufficient 
for the complete diagonalization of the system.

 The Hamiltonian theory of systems (\ref{OneDimSyst}) is connected, 
first of all, with local one-dimensional Poisson brackets of 
hydrodynamic type, introduced by B.A. Dubrovin and S.P. Novikov
(\cite{DubrovinNovikov1983}). The Dubrovin-Novikov bracket on the 
space of fields $\, (U^{1}(x), \dots, U^{N}(x)) \, $ has the form
\begin{multline}
\label{DNbr}
\{U^{\nu}(x), U^{\mu}(y)\} \,\,\, =  \\
= \,\,\, g^{\nu\mu}({\bf U}) \,\,
\delta^{\prime}(x-y) \,\,\, + \,\,\, 
b^{\nu\mu}_{\lambda}({\bf U}) \,\, U^{\lambda}_{x} \,\,
\delta (x-y) 
\end{multline}

 As was shown by B.A. Dubrovin and S.P. Novikov, the expression
(\ref{DNbr}) with a nondegenerate tensor
$\, g^{\nu\mu}({\bf U}) \, $ defines a Poisson bracket on the space
of fields $\, {\bf U}(x) \, $ if and only if:

1) The tensor  $\, g^{\nu\mu}({\bf U}) \, $ defines a symmetric 
pseudo-Riemannian metric of zero curvature with upper indices
on the space of parameters $\, (U^{1}, \dots, U^{N})$;

2) The values
$$\Gamma^{\nu}_{\mu\lambda} \,\,\, = \,\,\, - \,\, g_{\mu\sigma} 
\,\,  b^{\sigma\nu}_{\lambda} \,\,\, , $$
where
$\, g^{\nu\sigma}({\bf U}) \, g_{\sigma\mu}({\bf U}) = 
\delta^{\nu}_{\mu} \, $, represent the Christoffel symbols for
corresponding metric $\, g_{\nu\mu}({\bf U}) \, $.

 As follows from the above statements, any Dubrovin-Novikov bracket 
can be written in the canonical form(\cite{DubrovinNovikov1983})
\begin{equation}
\label{DNbrCanForm}
\{n^{\nu}(x), n^{\mu}(y)\} \,\,\, = \,\,\, \epsilon^{\nu} \,
\delta^{\nu\mu} \, \delta^{\prime}(x-y) \,\,\,\,\, , \,\,\,
\epsilon^{\nu} = \pm 1 
\end{equation}
after transition to the flat coordinates 
$\, n^{\nu} = n^{\nu}({\bf U})\, $ of the metric
$\, g_{\nu\mu}({\bf U})\, $. It is natural to introduce the group 
of point canonical transformations for the bracket 
(\ref{DNbrCanForm}), which, as is easy to see, coincides 
with the corresponding group $\, O (K, N-K) \, $.

 The functionals
$$N^{\nu} \,\, = \,\, \int_{-\infty}^{+\infty} 
n^{\nu}(x) \, d x $$
represent annihilators of the Dubrovin-Novikov bracket, 
while the functional
$$P \,\, = \,\, \int_{-\infty}^{+\infty} \, {1 \over 2} \,
\sum_{\nu=1}^{N} \, \epsilon^{\nu}  (n^{\nu})^{2} (x) \,\, d x $$
represents the momentum functional for the bracket (\ref{DNbr}).
 
 It is often also convenient to write the canonical form of 
the Dubrovin-Novikov bracket in a more general form
$$  \{n^{\nu}(x), n^{\mu}(y)\} \,\,\, = \,\,\, 
\eta^{\nu\mu} \, \delta^{\prime}(x-y) \,\,\, , $$
where $\, \eta^{\nu\mu} \, $ is an arbitrary constant 
(non-degenerate) symmetric matrix. It is easy to see that 
the group of (point) canonical transformations of 
the Poisson bracket is extended in this case to
$\, GL_{N} (\mathbb{R}) \, $.

  As one can see, if $\, N \, $ is even,
$\, N = 2 K $, and the metric $\, g_{\nu\mu}({\bf U})\, $ has 
the signature $\, (K, K) \, $, we can actually choose the 
flat coordinates
$\, (a^{1}, \dots , a^{K}, b^{1}, \dots , b^{K}) \, $
in such a way that the corresponding nonzero Poisson brackets 
take the form
$$  \{a^{\alpha}(x), b^{\beta}(y)\} \,\,\, = \,\,\, 
\delta^{\alpha\beta} \, \delta^{\prime}(x-y) $$

 In this case, making a real nonlocal transformation
$$\Phi^{\alpha} (x) \,\,\, = \,\,\, \int  a^{\alpha}(x) \, d x 
\,\,\, , $$
we get the Poisson bracket in canonical form
$$\{ b^{\alpha} (x) , b^{\beta} (y) \} \,\,\, = \,\,\, 0 \,\, ,
\quad \{ \Phi^{\alpha} (x) , \Phi^{\beta} (y) \} \,\,\, = \,\,\, 0 
\,\, ,  $$
$$\{ \Phi^{\alpha} (x) , b^{\beta} (y) \} \,\,\, = \,\,\, 
\delta^{\alpha\beta} \,\, \delta (x - y) $$

 To some extent, a special role is also played by the Poisson 
brackets, which are linear in the coordinates $\, {\bf U} (x) \, $:
\begin{multline}
\label{LinOneDimBr}
\{ U^{\nu} (x) , U^{\mu} (y) \} \,\,\, =   \\
 = \,\,\, \Big( \left( b^{\nu\mu}_{\lambda} \, + 
b^{\mu\nu}_{\lambda} \right) U^{\lambda} \, + \, g^{\nu\mu}_{0}
\Big) \,\, \delta^{\prime} (x - y) \,\,\, +  \\
+ \,\,\,  
b^{\nu\mu}_{\lambda} \, U^{\lambda}_{x} \,\, 
\delta (x - y) \,\,\, , 
\end{multline}
$$b^{\mu\nu}_{\lambda} \, = \, {\rm const} \quad , \,\,\,
g^{\nu\mu}_{0} \, = \, {\rm const} $$

 The coordinates $\, {\bf U} (x) \, $ in this case are, as a rule, 
naturally related to the problem under consideration, and the 
brackets (\ref{LinOneDimBr}) themselves are described by Lie-algebraic 
structures. In the case of one space variable, the classification 
of the corresponding Lie algebras, as well as admissible cocycles 
on them, was constructed in \cite{BalNov}, where, in particular, 
the connection of such brackets with the theory of Frobenius and 
quasi-Frobenius algebras was discovered.

\vspace{1mm}

 The bracket (\ref{DNbr}) also has two other important forms 
on the space of fields $\, {\bf U}(x) \, $. The first of them is 
the ``Liouville'' form
(\cite{DubrovinNovikov1983,DubrovinNovikov}), having the form
\begin{multline*}
\{U^{\nu}(x), U^{\mu}(y)\} \,\,\, =  \\
= \,\,\, \Big( \gamma^{\nu\mu}({\bf U}) \, + \,
\gamma^{\mu\nu}({\bf U}) \Big) \,\, \delta^{\prime}(x-y) 
\,\,\, + \,\,\,
{\partial \gamma^{\nu\mu} \over \partial U^{\lambda}} \, 
U^{\lambda}_{x} \,\, \delta (x-y)
\end{multline*}
for some functions $\, \gamma^{\nu\mu}({\bf U}) \, $.

 The ``Liouville'' form of the Dubrovin-Novikov bracket is also 
called physical and corresponds to the case when the integrals of 
coordinates $\, U^{\nu} \, $:
$$I^{\nu} \,\, = \,\, \int_{-\infty}^{+\infty} 
U^{\nu}(x) \, d x $$
commute with each other.

 Another important form of the Dubrovin-Novikov bracket is the 
diagonal form.  It corresponds to the case when the coordinates
$\, U^{\nu} \, $ are orthogonal coordinates for the metric 
$\, g_{\nu\mu}({\bf U}) \, $, and, accordingly, the tensor 
$\, g^{\nu\mu}({\bf U}) \, $ in the definition (\ref{DNbr}) has 
a diagonal form. This form of the Dubrovin-Novikov bracket is 
closely related to the theory of integrable systems of 
hydrodynamic type.

  S.P. Novikov hypothesized that all systems of hydrodynamic 
type that can be reduced to the form (\ref{DiagSyst}) and are 
Hamiltonian with respect to some bracket (\ref{DNbr}) are 
integrable. This hypothesis was proved by S.P. Tsarev in the 
work \cite{Tsarev1}, where a new method 
(``generalized hodograph method'') for integrating 
such systems was proposed.

 The construction of solutions of the system (\ref{DiagSyst}) 
by Tsarev's method consists in finding systems (commuting flows) 
that commute with it, which have the same (diagonal) form
with the characteristic velocities
$\, w^{\nu} ({\bf R}) \, $ satisfying the system of equations
\begin{equation}
\label{CommFlows}
{\partial_{\mu} w^{\nu} \over  w^{\mu} - w^{\nu}} 
\,\,\, = \,\,\, 
{\partial_{\mu} v^{\nu} \over  v^{\mu} - v^{\nu}}
\quad \quad , \quad \mu \neq \nu 
\end{equation}
($\partial_{\mu} \equiv \partial / \partial R^{\mu}$).

 Each of the solutions
$\, {\bf w} ({\bf R}) \, = \, 
(w^{1} ({\bf R}), \dots , w^{N} ({\bf R})) \, $
of the system (\ref{CommFlows}) generates a solution 
$\, {\bf R} (x,t) \, $ of the system (\ref{DiagSyst}), 
determined from the algebraic system
$$w^{\nu} ({\bf R}) \,\,\, = \,\,\, t \, v^{\nu} ({\bf R}) 
\,\,\, + \,\,\, x   \quad , \quad \quad
\nu \, = \, 1, \dots , N $$

 The system (\ref{CommFlows}) on the functions 
$\, w^{\nu} ({\bf R}) \, $ is an overdetermined system of 
linear equations with variable coefficients, the compatibility 
conditions for which have the form
\begin{equation}
\label{CompCond}
\partial_{\lambda} \left(
{\partial_{\mu} v^{\nu} \over  v^{\mu} - v^{\nu}} \right)
\,\,\, = \,\,\, \partial_{\mu} \left(
{\partial_{\lambda} v^{\nu} \over  v^{\lambda} - v^{\nu}} \right)
\end{equation}
$$\nu \neq \mu  \quad , \,\,\,  \mu \neq \lambda  
\quad , \,\,\,  \lambda \neq \nu $$

 As can be shown (see \cite{Tsarev1,Tsarev2}), the fulfillment 
of conditions (\ref{CompCond}) ensures in the general position 
the completeness of solutions of the system (\ref{CommFlows}), 
sufficient for the (local) solution of the general Cauchy problem 
for the corresponding system (\ref{DiagSyst}).

 As was shown in the work \cite{Tsarev1}, the requirement
that system (\ref{DiagSyst}) is Hamiltonian with respect to any 
Dubrovin-Novikov bracket entails the relations (\ref{CompCond})
and, thus, allows integration of this system by the Tsarev
method. In fact, the set of diagonal systems satisfying the 
condition (\ref{CompCond}) is much wider than the set of the 
same systems that are Hamiltonian with respect to a local 
Poisson bracket; therefore, all diagonal systems satisfying the 
conditions (\ref{CompCond}) were named by Tsarev semi-Hamiltonian.
As it turned out later, the class of ``semi-Hamiltonian'' 
systems also includes systems that are Hamiltonian with respect 
to generalizations of the Dubrovin-Novikov bracket - the weakly 
nonlocal Mokhov-Ferapontov bracket and the Ferapontov bracket. 
Let us give the appropriate definitions here:

\vspace{1mm}
 
 A Mokhov-Ferapontov bracket (\cite{MokhFer1}) on the space of
fields ${\bf U}(x)$ is a bracket that can be represented in the 
form
\begin{multline}
\label{MFbr}
\{ U^{\nu}(x), U^{\mu}(x) \} \,\,\, = \,\,\, 
g^{\nu\mu}({\bf U}) \,\, \delta^{\prime}(x-y) \,\,\, +  \\
+ \,\,\, b^{\nu\mu}_{\lambda}({\bf U}) \,\, 
U^{\lambda}_{x} \,\, \delta (x-y) \,\,\, + \,\,\,  
{1 \over 2} \,\, c \,\, U^{\nu}_{x} \,\, 
{\rm sgn} (x-y) \,\, U^{\mu}_{y}
\end{multline}

\vspace{1mm}

 A general Ferapontov bracket (\cite{Fer1,Fer2,Fer3,Fer4})
on the space of fields ${\bf U}(x)$ is a bracket of the form
\begin{multline}
\label{Fbr}
\{ U^{\nu}(x), U^{\mu}(y) \} \,\,\, =  \\
= \,\,\, g^{\nu\mu}({\bf U}) \,\, \delta^{\prime}(x-y) 
\,\,\, +  \,\,\, b^{\nu\mu}_{\lambda}({\bf U}) \,\,
U^{\lambda}_{x} \delta (x-y) \,\,\, +  \\
+ \,\,\, {1 \over 2} \,\, \sum_{k=1}^{g} \, e_{k}  \,\, 
w^{\nu}_{(k)\lambda}({\bf U}) \,
U^{\lambda}_{x} \,\,\,\, {\rm sgn} (x-y) \,\,\, 
w^{\mu}_{(k)\delta}({\bf U}) \, U^{\delta}_{y}
\end{multline}
$e_{k} = \pm 1$.

\vspace{1mm}

 Similarly to the previously formulated conditions on the 
functions $\, g^{\nu\mu}({\bf U}) \, $ and 
$\, b^{\nu\mu}_{\lambda}({\bf U})\, $
for the Dubrovin-Novikov bracket, one can formulate 
conditions on the coefficients in the expressions (\ref{MFbr}) 
and (\ref{Fbr}), under which they define the Poisson brackets 
on the space of fields $\, {\bf U}(x) \, $. Namely, the 
expression (\ref{MFbr}) with a nondegenerate tensor
$\, g^{\nu\mu}({\bf U}) \, $ defines a Poisson bracket on the
space of fields $\, {\bf U}(x) \, $ if and only if 
(\cite{MokhFer1}):

\vspace{1mm}

1) The tensor $\, g^{\nu\mu}({\bf U})\, $ defines a symmetric 
pseudo-Riemannian metric of constant curvature $c$ (with upper
indices) on the space of parameters
$\, (U^{1}, \dots, U^{N}) \, $;

2) The values
$$\Gamma^{\nu}_{\mu\gamma} \,\, = \,\, - \, g_{\mu\lambda} \,\,
b^{\lambda\nu}_{\gamma}$$
represent the Christoffel symbols for the corresponding metric
$\, g_{\nu\mu}({\bf U}) \, $.

\vspace{1mm}

 The expression (\ref{Fbr}) with a nondegenerate tensor
$\, g^{\nu\mu}({\bf U}) \, $ defines a Poisson bracket on the
space of fields $\, {\bf U}(x) \, $ if and only if (\cite{Fer1}):

1) The tensor $\, g^{\nu\mu}({\bf U})\, $ defines a symmetric 
pseudo-Riemannian metric with upper indices on the space of 
parameters $\, (U^{1}, \dots, U^{N}) \, $;

2) The values
$$\Gamma^{\nu}_{\mu\gamma} \,\, = \,\, - \, g_{\mu\lambda} \,\,
b^{\lambda\nu}_{\gamma}$$
represent the Christoffel symbols for the corresponding metric
$\, g_{\nu\mu}({\bf U}) \, $.

3) The affinors $\, w^{\nu}_{(k)\lambda}({\bf U}) \, $ and the 
curvature tensor of the metric 
$\, R^{\nu\tau}_{\mu\lambda}({\bf U}) \, $ satisfy the conditions
\begin{equation}
\label{metwein1}
g_{\nu\tau} w^{\tau}_{(k) \mu} = g_{\mu\tau} w^{\tau}_{(k) \nu},
\,\,\,\,\, \nabla_{\nu}  w^{\mu}_{(k) \lambda} =
\nabla_{\lambda}  w^{\mu}_{(k) \nu}
\end{equation}
\begin{equation}
\label{rtenw1}
R^{\nu\tau}_{\mu\lambda} \, = \, \sum_{k=1}^{g} e_{k}
\left(w^{\nu}_{(k) \mu} w^{\tau}_{(k) \lambda} -
w^{\tau}_{(k) \mu} w^{\nu}_{(k) \lambda} \right)
\end{equation}  
$$[w_{k}, w_{k^{\prime}}] = 0$$
(commutativity).

\vspace{1mm}

 E.V. Ferapontov also pointed out that the relations 
(\ref{metwein1})-(\ref{rtenw1}) represent the Gauss - Codazzi 
relations for a submanifold $\, {\cal M}^{N} \, $ with flat normal 
connection in a pseudo-Euclidean space $\, E^{N+g}$. In this 
situation, the tensor $\, g_{\nu\mu} \, $ plays the role of 
the first quadratic form on $\, {\cal M}^{N} \, $, and 
$\, w_{(k)} \, $ - the role of the Weingarten operators corresponding 
to ``parallel'' fields of unit normals $\, {\bf n}_{k} \, $
(\cite{Fer1,Fer2,Fer3,Fer4}. In addition, E.V. Ferapontov showed 
that the bracket (\ref{Fbr}) can be considered as a result of the 
Dirac restriction of the Dubrovin-Novikov bracket, defined in the 
space $\, E^{N+g} \, $, on the submanifold
$\, {\cal M}^{N} \, $ (\cite{Fer2}). 

\vspace{1mm}

 The canonical form of the Mokhov-Ferapontov bracket, written 
in the densities of annihilators, as well as the expression for 
the momentum functional for this bracket were proposed in
\cite{Pavlov1}. 

\vspace{1mm}

 Canonical forms (as well as ``canonical functionals'') of 
general weakly nonlocal Ferapontov brackets were investigated 
in \cite{PhysD}. By definition, the bracket (\ref{Fbr}) has 
the canonical form in the variables
$\, {\bf n} = {\bf n}({\bf U}) \, $ 
if the relations 
\begin{multline*}
\{n^{\nu}(x), n^{\mu}(y)\} \,\,\, =  \\
= \,\,\, \left( \epsilon^{\nu} \delta^{\nu\mu} - 
\sum_{k=0}^{g} \, e_{k} \, f^{\nu}_{(k)}({\bf n}) \, 
f^{\mu}_{(k)}({\bf n}) \right)
\delta^{\prime}(x-y) \,\,\, -  \\
\\ - \,\,\, \sum_{k=0}^{g} \, e_{k}
\left( f ^{\nu}_{(k)}({\bf n}) \right)_{x} 
f^{\mu}_{(k)}({\bf n}) \,\,
\delta (x-y) \,\,\, +  \\
+ \,\,\, {1 \over 2} \, \sum_{k=0}^{g} \, e_{k} \, 
\left( f^{\nu}_{(k)}({\bf n}) \right)_{x} \,\,
{\rm sgn} (x-y) \,\,
\left( f^{\mu}_{(k)}({\bf n}) \right)_{y} \,\,\,\,\, ,
\end{multline*}
($\epsilon^{\nu} = \pm 1$),
hold for some functions $\, f^{\nu}_{(k)}({\bf n}) \, $,
such that $\, f^{\nu}_{(k)}(0, \dots , 0) = 0 \, $.

 It can be noted here that the canonical form of the 
Ferapontov bracket in the general case is not uniquely 
defined. It is most natural to associate the canonical 
form of the bracket (\ref{Fbr}) with some fixed point
$\, {\bf U}_{0} \, $ on the variety of parameters
$\, {\bf U} \, $. It is natural to associate with the 
point $\, {\bf U}_{0} \, $ the space of ``loops'' starting 
and ending at the point $\, {\bf U}_{0} \, $, i.e. the
space $\, {\cal L}_{{\bf U}_{0}} \, $ of maps
$$\, \mathbb{R} \, \rightarrow \, \{{\bf U}\} \,\,\, , $$
such that
$${\bf U} (x) \,\,\, \rightarrow \,\,\, {\bf U}_{0}
\quad , \,\,\, x \rightarrow \pm \infty $$

 From the geometric point of view, the construction of the 
canonical coordinates corresponding to the point  
$\, {\bf U}_{0} \, $ is related to the above-mentioned 
embedding of the variety of parameters $\, {\cal M}^{N} \, $ 
into pseudo-Euclidean space $\, E^{N+g} \, $, defined by the 
bracket (\ref{Fbr}). Namely, consider the corresponding 
embedding
$$\, {\cal M}^{N} \,\,\, \rightarrow \,\,\, E^{N+g} $$
and choose a (pseudo) Euclidean coordinate system in
$\, E^{N+g} \, $ starting at the point
$\, {\bf U}_{0} \in {\cal M}^{N} \subset E^{N+g} \, $
so that the first $\, N \, $ coordinates are tangent to the 
submanifold $\, {\cal M}^{N} \, $ at the point 
$\, {\bf U}_{0} \, $, and the remaining $\, g \, $ 
are orthogonal to $\, {\cal M}^{N} \, $ at this point.
The restriction of the first $\, N \, $ coordinates in 
$\, E^{N+g} \, $ to $\, {\cal M}^{N} \, $ define then
the set of canonical coordinates $\, n^{\nu} ({\bf U}) \, $,
corresponding to the point $\, {\bf U}_{0} \, $. The 
restriction of the remaining $\, g \, $ coordinates to the 
submanifold $\, {\cal M}^{N} \, $ defines $\, g \, $ 
additional functions $\, h^{k} ({\bf U}) \, $, also playing 
an important role.

 The values of the (vector) functions 
$\, {\bf f}_{(k)} ({\bf n}) \, $ in the above canonical form 
coincide with the values of the projections of the basic 
parallel normal fields to $\, {\cal M}^{N} \, $ in the space 
$\, E^{N+g} \, $ onto the plane tangent to $\, {\cal M}^{N} \, $ 
at the point $\, {\bf U}_{0} \, $. In particular, for the 
Mokhov-Ferapontov bracket, corresponding to the case of a 
(pseudo) sphere in $\, E^{N+1} \, $, we have
$\, f^{\nu} \, = \, \sqrt{|c|} \, n^{\nu} \, $.

 As can be shown (see \cite{PhysD}), the corresponding
functionals
$$N^{\nu} \,\,\, = \,\,\, \int_{-\infty}^{+\infty}
n^{\nu} (x) \,\, d x $$
define the annihilators of the bracket (\ref{Fbr}) on the 
space $\, {\cal L}_{{\bf U}_{0}}$. The functionals
$$H^{k} \,\,\, = \,\,\, \int_{-\infty}^{+\infty}
h^{k} (x) \,\, d x $$
represent the Hamiltonians, generating the flows
$$U^{\nu}_{t} \,\,\, = \,\,\, w^{\nu}_{(k)\mu}({\bf U}) \,\,
U^{\mu}_{x} $$
on the space $\, {\cal L}_{{\bf U}_{0}}$ according to the
bracket (\ref{Fbr}).

 Every bracket (\ref{Fbr}) with a nondegenerate tensor
$\, g^{\nu\mu}({\bf U)} \, $ also has a local momentum functional 
$\, P \, $ on the spaces $\, {\cal L}_{{\bf U}_{0}}$
having the form
$$P \,\,\, = \,\,\, {1 \over 2} \int_{-\infty}^{+\infty} \left(
\sum_{\nu=1}^{N} \epsilon^{\nu} n^{\nu} n^{\nu} \,\, + \,\,   
\sum_{k=1}^{g} e_{k} \, h^{k} h^{k} \right) d x  \quad , $$
where the functions $\, {\bf n} ({\bf U}) \, $ and 
$\, {\bf h} ({\bf U}) \, $ correspond to the point 
$\, {\bf U}_{0} \, $ (\cite{IJMMS2}).

 In \cite{PhysD}, there was also proposed a Liouville 
(physical) form of a general Ferapontov bracket, 
which has the form
\begin{multline*}
\{U^{\nu}(x),U^{\mu}(y)\} \,\,\, =  \\
= \,\,\, \left(\gamma^{\nu\mu} \, + \, \gamma^{\mu\nu}
\, - \, \sum_{k=1}^{g} e_{k} f^{\nu}_{(k)} f^{\mu}_{(k)}
\right) \delta^{\prime}(x-y) \,\,\, +  \\
+ \,\,\, \left(
{\partial \gamma^{\nu\mu} \over \partial U^{\lambda}} \,
U^{\lambda}_{x} \,\, - \,\, 
\sum_{k=1}^{g} \, e_{k} \, (f^{\nu}_{(k)})_{x} \,
f^{\mu}_{(k)}\right) \delta (x-y) \,\,\, +  \\
+ \,\,\, {1 \over 2} \, 
\sum_{k=1}^{g} \, e_{k} \, (f^{\nu}_{(k)})_{x} \,\,\, 
{\rm sgn} (x-y) \,\,\, (f^{\mu}_{(k)})_{y}
\end{multline*}
with some functions $\, \gamma^{\nu\mu}({\bf U}) \, $ and 
$\, f^{\nu}_{(k)}({\bf U})\, $.

 The bracket (\ref{Fbr}) has the physical form in coordinates
$\, U^{\mu} \, $ if and only if the integrals
$$J^{\nu} \, = \, \int_{-\infty}^{+\infty} U^{\nu}(x) \,\, d x$$
generate a set of local commuting flows according to the bracket 
(\ref{Fbr}).

 The Ferapontov bracket (\ref{Fbr}) is the most general 
(one-dimensional) weakly nonlocal hydrodynamic bracket. 
On the other hand, the Ferapontov brackets, as a rule, are 
associated only with integrable systems of hydrodynamic type, 
since, in reality, they contain integrable structures within 
themselves. According to the hypothesis of E.V. Ferapontov 
(\cite{Fer4}), all diagonalizable semi-Hamiltonian systems 
are actually Hamiltonian with respect to brackets (\ref{Fbr}), 
if we admit the presence of an infinite number of terms in 
their non-local part. This hypothesis was, in particular, 
confirmed in the work \cite{BogdanovFerapontov} for a fairly 
wide class of semi-Hamiltonian systems. This hypothesis was 
also considered in the works \cite{ZakharovDifGeom} and 
\cite{MokhovSemiHam}. It must be said, however, that in the 
most general formulation a rigorous proof of this hypothesis 
has not yet been obtained.

 At the same time, the brackets (\ref{DNbr}) and (\ref{MFbr}) 
are associated with a much broader class of systems that do not 
imply integrability, and are, in this sense, more common in 
real applications.

 Both in the case of the Dubrovin-Novikov bracket and in the 
case of the weakly nonlocal brackets of Mokhov-Ferapontov and 
Ferapontov, the diagonal form of the metric $\, g_{\nu\mu} \, $
is most closely related to the theory of integration of systems
(\ref{OneDimSyst}). With all this, the problem of diagonal 
coordinates for flat metrics has been, as is known, a classical 
problem in differential geometry for many years. It should be 
noted here that recently very important new achievements 
(\cite{Zakh1,Krichev1}) related to the inverse scattering 
problem method have been obtained in solving this problem.
It can be noted that the theory of Poisson brackets also 
served as a stimulating factor in considering this classical 
problem.

\vspace{1mm}

 It must be said, however, that the diagonal form of the 
Poisson brackets is, certainly, not the only one important 
when considering the integrability of systems of hydrodynamic 
type. So, for example, the canonical coordinates of the 
brackets play the most important role when considering 
bi-Hamiltonian structures (see \cite{Magri}), which are 
present in the overwhelming majority of integrable 
hierarchies. Recall here that the basis of a bi-Hamiltonian 
structure is represented by a pair of compatible Poisson 
brackets, $\, \{ \dots , \dots \}_{1} \, $ and
$\, \{ \dots , \dots \}_{2} \, $, i.e. brackets, any linear 
combination of which
\begin{equation}
\label{PencilBr}
\{ \dots , \dots \}_{1} \,\,\, + \,\,\, \lambda \,
\{ \dots , \dots \}_{2} 
\end{equation}
also defines a Poisson bracket. The construction of 
bi-Hamiltonian hierarchies, as a rule, begins with 
annihilators or some canonical functionals associated with 
one of the Hamiltonian structures and playing the role of 
``ancestors'' of the corresponding hierarchies. As we saw above, 
in the case of brackets of hydrodynamic type, such functionals 
are related precisely to the canonical coordinates of the bracket.
It can be added that the canonical coordinates
$\, n^{\nu} (\lambda) \, $ for the entire pencil (\ref{PencilBr})
represent in this case generating functions for all densities 
of the Hamiltonians of the hierarchy under consideration that 
generate higher flows.

 It is also impossible not to mention here the important role 
of compatible local Hamiltonian structures for the quantum field 
theory discovered by B.A. Dubrovin in the 1990s. Namely, as 
was shown by B.A. Dubrovin, special pairs of compatible 
Dubrovin-Novikov brackets give a classification of topological 
field theories that arise as approximations in various field 
theory models. Thus, a pair of matched Dubrovin-Novikov brackets 
underlies the definition of Frobenius manifolds 
(\cite{Dubrov1,Dubrov3}), parametrizing solutions of the WDVV 
equation describing topological theories. Moreover, 
the further development of the theory of Frobenius manifolds 
was the theory of ``weakly dispersive'' deformations of such 
Hamiltonian structures, which give corrections to topological 
theories (\cite{Dubrov4,Dubrov5}).

\vspace{1mm}

 In addition, certainly, in many examples the most important 
role is played by the coordinates at which the Poisson bracket 
becomes linear in the fields, where the Lie algebraic aspects 
of the problem under consideration are clearly manifested.

\vspace{1mm}

 Another important generalization of local brackets of 
hydrodynamic type are inhomogeneous Poisson brackets, which in 
the general case have the form
\begin{multline}
\label{NeodnMnogSkob}
\{ U^{\nu} ({\bf x}) , U^{\mu} ({\bf y}) \} \,\,\, =  \,\,\, 
g^{\nu\mu i} \left( {\bf U} ({\bf x}) \right) \,\,
\delta_{x^{i}} ({\bf x} - {\bf y}) \,\,\,  +  \\
+ \,\,\,  \left[ 
b^{\nu\mu i}_{\lambda} \left( {\bf U} ({\bf x}) \right) 
U^{\lambda}_{x^{i}} \,\, + \,\, h^{\nu\mu} ({\bf U}) \right] \,
\delta ({\bf x} - {\bf y}) 
\end{multline}

 The brackets (\ref{NeodnMnogSkob}) represent actually pairs of 
compatible brackets, since both their hydrodynamic part
(i.e. (\ref{GenBr})), and the finite-dimensional part 
$\, h^{\nu\mu} ({\bf U}) \, \delta ({\bf x} - {\bf y}) $,
define independent Poisson brackets on the space of fields
$\, {\bf U} ({\bf x}) \, $, compatible with each other.
It can be seen, in addition, that the tensor
$\, h^{\nu\mu} ({\bf U}) \, $ must also define a simple 
finite-dimensional bracket on a space with coordinates
$\, {\bf U} \, $. 

 Brackets (\ref{NeodnMnogSkob}), as it is easy to see, are 
associated primarily with inhomogeneous systems of hydrodynamic 
type
$$U^{\nu}_{t} \,\,\, = \,\,\, V^{\nu i}_{\mu} 
\left( {\bf U} \right) \, U^{\mu}_{x^{i}} \,\, + \,\, 
f^{\nu} ({\bf U}) $$
on the space of fields $\, {\bf U} ({\bf x}) \, $.

 The theory of brackets (\ref{NeodnMnogSkob}) is also quite 
non-trivial and contains many interesting geometric and algebraic 
structures. Thus, in particular, for one-dimensional Poisson 
brackets
\begin{multline}
\label{NeodnOdnomernSkob}
\{ U^{\nu} (x) , U^{\mu} (y) \} \,\,\, =  \,\,\, 
g^{\nu\mu} \left( {\bf U} (x) \right) \,\,
\delta^{\prime} (x - y) \,\,\,  +  \\
+ \,\,\,  \left[ 
b^{\nu\mu}_{\lambda} \left( {\bf U} \right) 
U^{\lambda}_{x} \,\, + \,\, h^{\nu\mu} ({\bf U}) \right] \,
\delta (x - y) 
\end{multline}
the following statement is true (\cite{DubrNovMnogomernSkob}):

 If the metric $\, g^{\nu\mu} \left( {\bf U} \right) \, $ 
is nondegenerate, then, after a coordinate transformation 
$\, {\bf U} = {\bf U} ({\bf n}) \, $, the bracket  
(\ref{NeodnOdnomernSkob}) can be written in the form
\begin{multline*}
\{ n^{\nu} (x) , n^{\mu} (y) \} \,\,\, =  \,\,\, 
\eta^{\nu\mu} \,\, \delta^{\prime} (x - y) \,\,\,  +  \\
+ \,\,\,  \left[ 
c^{\nu\mu}_{\lambda} \, n^{\lambda} \,\, + \,\, 
d^{\nu\mu} \right] \, \delta (x - y) \,\,\, ,
\end{multline*}
where the coefficients  $\, \eta^{\nu\mu} \, $,
$\, c^{\nu\mu}_{\lambda} \, $, $\, d^{\nu\mu} \, $
are constant, $\, c^{\nu\mu}_{\lambda} \, $ represent the 
structure constants of a semisimple Lie algebra with Killing 
metric $\, \eta^{\nu\mu} \, $, and
$\, d^{\nu\mu} \, = \, - \, d^{\mu\nu} \, $ is an arbitrary 
cocycle on this algebra.

 An important example of a system associated with the bracket 
(\ref{NeodnOdnomernSkob}) is the $n$-waves problem
\begin{equation}
\label{nwaves}
M_{t} \,\, - \,\, \varphi \left( M_{x} \right) 
\,\,\, = \,\,\, \left[ M , \varphi \left( M \right) 
\right] \,\,\, , 
\end{equation}
where $\, M = \left( M_{ij} \right) \, $ is an 
$ \, (n \times n)$ - matrix with zero trace
(possibly with additional symmetries),
$\, \varphi \left( M \right) \, = \,
\left( \lambda_{ij} M_{ij} \right) \, $,
which is Hamiltonian with respect to a bracket of the form
(\ref{NeodnOdnomernSkob}) with $\, d^{\nu\mu} = 0 \, $,
with a quadratic Hamiltonian. As is well known 
(see \cite{ZakhManNovPit}), the system (\ref{nwaves}) is 
integrable by the inverse scattering problem method for 
$\, \lambda_{ij} \, = \, (a_{i} - a_{j}) / (b_{i} - b_{j}) \, $.

 We also note here that Hamiltonian structures 
(\ref{NeodnOdnomernSkob}) have further generalizations. 
So, in particular, in the work \cite{MokhFerFAA} inhomogeneous 
Poisson brackets corresponding to metrics of constant curvature 
(Mokhov - Ferapontov brackets) were considered.

\vspace{1mm}

 In conclusion of this chapter, it can be noted that the 
brackets (\ref{DNbr}) as well as (\ref{MFbr}) - (\ref{Fbr}), 
(\ref{NeodnOdnomernSkob}), represent special classes of more 
general local one-dimensional field-theoretic brackets
\begin{equation}
\label{localbr}
\{\varphi^{i}(x) , \varphi^{j}(y)\} \,\, = \,\, \sum_{k \geq 0}
B^{ij}_{(k)} ({\varphi}, {\varphi}_{x}, \dots)
\, \delta^{(k)}(x-y)
\end{equation}
and weakly non-local brackets
$$\{\varphi^{i}(x) , \varphi^{j}(y)\} \,\, = \,\, \sum_{k \geq 0}
B^{ij}_{(k)} ({\varphi}, {\varphi}_{x}, \dots)
\, \delta^{(k)}(x-y) \,\, + $$
\begin{equation}
\label{nonlocbr}
+ \,\, {1 \over 2} \, \sum_{k,s=1}^{g} \kappa_{ks} \,\,
S^{i}_{(k)} ({\varphi}, {\varphi}_{x}, \dots) \,\,
{\rm sgn} (x-y) \,\,
S^{j}_{(s)} ({\varphi}, {\varphi}_{y}, \dots) \, ,
\end{equation}
where $\, \kappa_{ks} \, $ is some arbitrary (constant) quadratic 
form, $\, i = 1, \dots, n \, $, 
$\,\, {\varphi}(x) = (\varphi^{1}(x), \dots, \varphi^{n}(x)) \, $.

 As in the case of brackets of hydrodynamic type, brackets
(\ref{localbr}) can be associated with a very wide variety 
of systems of evolutionary type, while brackets
(\ref{nonlocbr}) appear, as a rule, in integrable 
hierarchies (see \cite{EnOrRub,PhysD}).

\section{Hamiltonian structures and the theory of slow modulations}
\setcounter{equation}{0}

 The development of the theory of integration of one-dimensional 
systems of hydrodynamic type actually took place in close 
connection with the development of the theory of slow modulations 
(Whitham's method) of multiphase solutions of integrable 
hierarchies. As was first shown by G. Whitham (\cite{Whitham}), 
the averaged equations describing the evolution of slowly modulated 
parameters of single-phase solutions of the KdV equation are a 
hydrodynamic-type system reduced to a diagonal form.
As was later shown in \cite{FFM}, this remarkable property is also 
inherent in the equations of slow modulations (Whitham's equations) 
for multiphase KdV solutions. It can be immediately noted here that 
the construction of multiphase solutions for integrable hierarchies 
is closely related to the methods of algebraic geometry in the theory 
of the inverse scattering problem (see 
\cite{novikov1994,DubrNov1974A,DubrNov1974B,ItsMatveev1,
ItsMatveev2,DubrMatvNov,DubrKrichNov}). As a consequence of this, 
the methods of the theory of slow modulations for integrable systems 
are also closely related to algebraic-geometric constructions.
In particular, it was the methods of algebraic geometry that made 
it possible to establish the diagonalizability of Whitham's equations 
for multiphase solutions of KdV in the paper \cite{FFM}.
Being a universal approach in the theory of integrable systems, 
the methods of algebraic geometry make it possible actually to 
establish a similar fact for most hierarchies integrable by the 
method of the inverse scattering problem. Thus, systems of equations 
for slow modulations for integrable hierarchies represent the most 
important class of diagonalizable systems of hydrodynamic type.

 As we have already said, integration of diagonal systems of 
hydrodynamic type usually requires their Hamiltonian nature. 
In some cases, the Hamiltonian nature of systems of equations 
can be related to their Lagrangian formalism, which, as a rule, 
allows us to determine the corresponding Hamiltonian structure 
in canonical form. Along with the procedure for constructing 
equations of slow modulations, Whitham (\cite{Whitham}) also 
proposed a procedure for averaging local Lagrangians 
(also in the presence of ``pseudophases'') and obtaining a local 
Lagrangian formalism for the ``averaged system''. It must be said, 
however, that not all interesting systems of evolutionary type 
\begin{equation}
\label{ldsyst}
\varphi^{i}_{t} \,\, = \,\,
Q^{i} ({\varphi}, {\varphi}_{x}, \dots) 
\end{equation}
have local Lagrangian structures, and the most common for such 
systems is, as a rule, the presence of a more general Hamiltonian 
formalism with the Poisson bracket (\ref{localbr}) 
(or (\ref{nonlocbr})) and the Hamiltonian
\begin{equation}
\label{InHamilt}
H \,\, = \,\, \int_{-\infty}^{+\infty}
h ({\varphi}, {\varphi}_{x}, \dots) \, dx 
\end{equation}

 To construct Hamiltonian structures for the equations of slow 
modulations in the general case, B.A. Dubrovin and S.P. Novikov
(\cite{DubrovinNovikov1983,DubrovinNovikov}) proposed a procedure 
of ``averaging'' of local Poisson brackets (\ref{localbr}),
which gives a bracket of the form (\ref{DNbr}) for the Whitham 
system. Let us give here a brief description of this procedure.

 Namely, in Whitham's method, we assume that the system 
(\ref{ldsyst}) has a finite-parameter family of quasiperiodic 
solutions
\begin{equation}
\label{mphasesol}
{\varphi}^{i}(x,t) \,\,\, = \,\,\,
\Phi^{i} \left({\bf k}({\bf U}) \, x \, + \,
\bm{\omega}({\bf U}) \, t \, + \, \bm{\theta}_{0},
\, {\bf U} \right) \,\,\, , 
\end{equation}
where 
$\bm{\theta} \, = \, (\theta^{1}, \dots, \theta^{m})$,
$\, {\bf k} ({\bf U}) \, = \, 
(k^{1}({\bf U}), \dots, k^{m}({\bf U}))$, 
$\bm{\omega} ({\bf U}) \, = \, 
(\omega^{1}({\bf U}), \dots, \omega^{m}({\bf U}))$
and $\, \Phi^{i} ({\theta}, {\bf U})$ define a family of 
$2\pi$-periodic in all $\theta^{\alpha}$ functions that depend 
on additional parameters ${\bf U} \, = \, (U^{1}, \dots, U^{N})$.

 In the Whitham method, we stretch both the 
coordinates $x$ and $t$: $\, X = \epsilon \, x$, 
$\, T = \epsilon \, t \, $, ($\epsilon \rightarrow 0$),  and 
consider the parameters $\, {\bf U} \, $ as functions 
of ``slow'' coordinates $X$ and $T$.

 The method of B.A. Dubrovin and S.P. Novikov is based on the 
existence of $N$ (equal to the number of parameters $U^{\nu}$ on the 
family of $m$-phase solutions of (\ref{ldsyst})) local integrals
$$I^{\nu} \,\, = \,\, 
\int {\cal P}^{\nu}(\varphi,\varphi_{x},\dots) \, dx 
\,\,\, , $$
commuting with the Hamiltonian and with each other
\begin{equation}
\label{invv}
\{I^{\nu} , H\} = 0 \,\,\,\,\, , \,\,\,\,\,
\{I^{\nu} , I^{\mu}\} = 0 \,\,\, ,
\end{equation}
and can be described as follows:

 Let us calculate the pairwise Poisson brackets of the 
densities ${\cal P}^{\nu}$, having the form
 $$\{{\cal P}^{\nu}(x), {\cal P}^{\mu}(y)\} =
\sum_{k\geq 0} \,
A^{\nu\mu}_{k}(\varphi,\varphi_{x},\dots) \,
\delta^{(k)}(x-y) \,\,\, , $$
where
$$A^{\nu\mu}_{0}(\varphi,\varphi_{x},\dots) \equiv
\partial_{x} Q^{\nu\mu}(\varphi,\varphi_{x},\dots) $$
according to (\ref{invv}). The corresponding Dubrovin - Novikov 
bracket on the space of functions $U(X)$ has the form:
\begin{multline}
\label{dubrnovb}
\{U^{\nu}(X), U^{\mu}(Y)\} \,\,\, =   \\
= \,\, \langle A^{\nu\mu}_{1}\rangle (U) \,\, 
\delta^{\prime}(X-Y) \,\,\, +  \,\,\,
{\partial \langle Q^{\nu\mu} \rangle \over 
\partial U^{\lambda}} \, U^{\lambda}_{X} \,\, \delta (X-Y)
\end{multline}
where $\langle \dots \rangle$ means the averaging over the 
family of $m$-phase solutions of (\ref{ldsyst}) given by the 
formula
$$\langle F \rangle \,\, = \,\, 
{1 \over (2\pi)^{m}} \int_{0}^{2\pi}\!\!\dots\int_{0}^{2\pi}
F (\Phi, k^{\alpha}(U) \Phi_{\theta^{\alpha}}, \dots) \,
d^{m}\theta $$

 In this case, we choose the parameters $U^{\nu}$ in the form
$\, U^{\nu} = \langle P^{\nu} \rangle \, $
on the corresponding solutions of the family under consideration.

 It can be seen that the Dubrovin and Novikov method in the 
described form is associated with the conservative form of the 
system of equations for slow modulations, namely, with the form
\begin{equation}
\label{ConsWhith}
U^{\nu}_{T} \,\,\, = \,\,\, \langle Q^{\nu} \rangle_{X} 
\quad , \quad \quad  \nu = 1, \dots , N \,\,\, , 
\end{equation}
where
$${\cal P}^{\nu}_{t} \,\,\, \equiv \,\,\, 
{\partial \over \partial x} \,\, 
Q^{\nu} (\varphi, \varphi_{x}, \dots ) $$
by virtue of the system (\ref{ldsyst}). It can also be shown that 
the bracket (\ref{dubrnovb}) generally has a Liouville form.
The integrals of the coordinates $\, U^{\nu} (X) \, $
give in this case the set of $\, N \, $ commuting integrals of the 
system (\ref{ConsWhith}), and the density of the Hamiltonian for the 
system (\ref{ConsWhith}) is given by the averaged density of the 
Hamiltonian of the original system $\, \langle h \rangle (X) \, $.
 
 It can also be noted here that all integrals $\, I^{\mu} \, $
generate commuting flows to the system (\ref{ldsyst}) according
to the corresponding bracket (\ref{localbr}), so that, along with 
time $\, t \, $, we can consider also the dependence of all 
solutions on additional times $\, t^{\mu} \, $
($\mu = 1, \dots , N$). In this case we have also the relations
$${\cal P}^{\nu}_{t^{\mu}} \,\,\, \equiv \,\,\, 
{\partial \over \partial x} \,\, 
Q^{\nu\mu} (\varphi, \varphi_{x}, \dots ) $$
for some functions
$\, Q^{\nu\mu} (\varphi, \varphi_{x}, \dots ) \, $,
and the corresponding family of solutions (\ref{mphasesol})
also represents a family of $m$ - phase solutions for
commuting flows with some other values of 
$\, \bm{\omega}^{\mu} ({\bf U}) \, = \, 
(\omega^{1 \mu}({\bf U}), \dots, \omega^{m \mu}({\bf U})) \, $.

 Whitham systems for the commuting flows 
$$U^{\nu}_{T^{\mu}} \,\,\, = \,\,\, 
\langle Q^{\nu\mu} \rangle_{X} $$
represent commuting flows for the system (\ref{ConsWhith}) 
and are Hamiltonian with respect to the bracket (\ref{dubrnovb}) 
with the Hamiltonians
$$H^{\mu} \,\,\, = \,\,\, \int_{-\infty}^{+\infty}
U^{\mu} (X) \, d X $$

 In general, the justification of the Dubrovin - Novikov method 
requires certain conditions of completeness and regularity of the 
corresponding family of $m$ - phase solutions of (\ref{ldsyst}) 
(see e.g. \cite{SIGMA}).

 Note here that the Dubrovin - Novikov procedure also admits 
a generalization to the case of weakly nonlocal Hamiltonian 
structures, which makes it possible to obtain the brackets 
(\ref{Fbr}) for the Whitham system by averaging the brackets 
(\ref{nonlocbr}) for the original system (\cite{IJMMS1}). 
On the whole, this procedure is quite convenient both for the 
case of integrable hierarchies and in a more general situation.

 The representaion of averaged Poisson brackets in diagonal form 
for integrable hierarchies, as well as the representaion of the 
corresponding Whitham systems in this form, is associated with 
algebraic-geometric methods of the inverse scattering problem 
(see e.g. \cite{DubrovinNovikov1983,DubrovinFAA,Tsarev2,Alekseev}).
In more detail, as in the case of diagonalization of the Whitham 
systems themselves, the diagonal coordinates for the averaged 
Poisson brackets for such hierarchies are related to the branch 
points of the Riemann surfaces that determine the corresponding 
$m$ - phase solutions of the original system. As noted above, the 
diagonal form of such brackets is most closely related to the 
procedure for integrating the corresponding systems of hydrodynamic 
type. It can also be noted that the construction of the most 
interesting solutions to averaged equations is also 
based on algebraic-geometric methods (see e.g.
\cite{GurevichPitaevskii1,GurevichPitaevskii2,AvKrichNov,
Krichever1988,Potemin2,GurKrEl1,GurKrEl2,Tian,TGrava1,
TGrava2,TGrava3,TGravaTian}).

\vspace{1mm}

 Let us also focus here on the canonical forms of the averaged 
brackets. 

 One of the features of the canonical form of averaged 
Poisson brackets (see e.g. \cite{Whitham,NovMal1993,MalPav,SIGMA})
is that all the functions $\, k^{\alpha} ({\bf U}) \, $ are part of 
the canonical coordinates of the averaged bracket and thus represent 
a part of the flat coordinates for the corresponding metric
$\, g^{\nu\mu} ({\bf U}) \, $ satisfying the relations
\begin{equation}
\label{kalphakbeta}
\{ k^{\alpha} (X) , k^{\beta} (Y) \} \,\,\, = \,\,\, 0 
\end{equation}

 In addition, it can be shown that the Poisson brackets of
$\, k^{\alpha} (X) \, $ with the densities 
$\, U^{\nu} (Y) \, $ can be written as
\begin{equation}
\label{kalphaUnu}
\{ k^{\alpha} (X) , U^{\nu} (Y) \} \,\,\, = 
\Big( \omega^{\alpha\nu} (X) \, \delta (X-Y) \Big)_{X} 
\end{equation}

  As was shown in \cite{MinimalSet}, the relations
(\ref{kalphakbeta}) - (\ref{kalphaUnu}) also allow us to 
modify the Dubrovin - Novikov procedure by reducing the 
required number of commuting integrals of the original 
system to $\, N - m \, $ and representing the averaged 
bracket in coordinates
$\, (k^{1}, \dots, k^{m}, \, U^{1}, \dots, U^{N-m}) \, $
in the form
(\ref{kalphakbeta}) - (\ref{kalphaUnu}) and 
\begin{multline}
\label{ModDubrNovB}
\{U^{\nu}(X), U^{\mu}(Y)\} \,\,\, = \,\,\, 
\langle A^{\nu\mu}_{1}\rangle (U) \,\, 
\delta^{\prime}(X-Y) \,\,\, +  \\
+ \,\,\, {\partial \langle Q^{\nu\mu} \rangle \over 
\partial U^{\lambda}} \, U^{\lambda}_{X} \,\, 
\delta (X-Y) \,\,\, , \quad 
\nu , \, \mu \, = \, 1, \dots , N - m
\end{multline}

 The representation of the averaged bracket in the form 
(\ref{kalphakbeta}) - (\ref{ModDubrNovB}) is actually connected 
with another important (equivalent) representation of the system 
of slow modulation equations, separating the phase evolution 
equations
$$S^{\alpha}_{T} \,\,\, = \,\,\, 
\omega^{\alpha} ({\bf S}_{X}, {\bf U}) $$
($S^{\alpha}_{X} \equiv k^{\alpha}$) and the transport equations,
which can be written as the balance equations for a part of the 
conservation laws
$$U^{\nu}_{T} \,\,\, = \,\,\, \langle Q^{\nu} \rangle_{X} 
\quad , \quad \quad  \nu = 1, \dots , N - m $$

 The representation of a bracket in the form
(\ref{kalphakbeta}) - (\ref{ModDubrNovB}) can also be called 
a mixed representation of the averaged bracket, which partially 
uses canonical and partially - Liouville variables. In the 
one-dimensional case, as already mentioned, the bracket can be 
completely reduced to a constant form, thus, in this case, 
we can choose the variables
$\, {\bf J} = (J_{1}, \dots , J_{m}) \, $,
$\, {\bf n} = (n^{1}, \dots , n^{N-2m}) \, $,
such that
$$\{k^{\alpha}(X) , J_{\beta}(Y) \} \,\, = \,\,
\delta^{\alpha}_{\beta} \, \delta^{\prime} (X-Y) \,\,\, , $$
$$\{ n^{q}(X) , n^{p}(Y) \} \,\, = \,\, \epsilon^{q} \, 
\delta^{qp} \,\, \delta^{\prime} (X-Y) $$
(the rest of the brackets are zero).

 Approaching the usual terminology in the theory of Poisson 
brackets, the relations
$$\{S^{\alpha}(X) , J_{\beta}(Y) \} \,\, = \,\,
\delta^{\alpha}_{\beta} \, \delta (X-Y) \,\,\, , $$
$$\{ n^{q}(X) , n^{p}(Y) \} \,\, = \,\, \epsilon^{q} \, 
\delta^{qp} \,\, \delta^{\prime} (X-Y) $$
could also be called the pseudo-canonical form of
brackets of hydrodynamic type.

 As was also noted in \cite{MinimalSet}, using the modified 
Dubrovin-Novikov procedure and representing the averaged brackets 
in the form (\ref{kalphakbeta}) - (\ref{ModDubrNovB}) is 
especially convenient in the case of several spatial variables.
In the next chapter, we will discuss in more detail 
the ``multi-dimensional'' Poisson brackets.

\section{Multi-dimensional hydrodynamic brackets and their 
generalizations}
\setcounter{equation}{0}

 We now turn to more general multi-dimensional brackets of the form 
(\ref{GenBr}):
$$\{ U^{\nu} ({\bf x}) , U^{\mu} ({\bf y}) \} \,\,\, =   $$
$$ = \,\,\, g^{\nu\mu i} \left( {\bf U} ({\bf x}) \right) \,\,
\delta_{x^{i}} ({\bf x} - {\bf y}) \,\, + \,\, 
b^{\nu\mu i}_{\lambda} \left( {\bf U} ({\bf x}) \right) 
U^{\lambda}_{x^{i}} \,\, \delta ({\bf x} - {\bf y}) \,\,\, , $$
the most complete theory of which was constructed in the works
\cite{DubrNovMnogomernSkob,MokhovMnogomernSkob1,MokhovMnogomernSkob2}.

 We consider here the case of non-degenerate brackets (\ref{GenBr}),
namely, we require that all tensors
$\, g^{\nu\mu i} ( {\bf U} ) \, $ are non-degenerate
$${\rm det} \,\, g^{\nu\mu i} ( {\bf U} ) \,\, \neq \,\, 0 \quad ,
\,\,\, i = 1, \dots , n $$
 
 As in the one-dimensional case, here we can also assert that all 
tensors $\, g^{\nu\mu i} ( {\bf U} ) \, $ represent flat metrics on the 
space of parameters $\, {\bf U} \, $, while the values
$$\Gamma^{\nu}_{\mu\lambda i} 
\,\,\, = \,\,\, - \,\, g_{\mu\sigma i} 
\,\,  b^{\sigma\nu i}_{\lambda} $$
give the corresponding Christoffel symbols. In general, to determine 
a correct Poisson bracket, the coefficients  
$\, g^{\nu\mu i} ( {\bf U} ) \, $ and
$\, b^{\nu\mu i}_{\lambda} ( {\bf U} ) \, $ 
must satisfy a number of nontrivial relations
(see \cite{DubrNovMnogomernSkob,MokhovMnogomernSkob2}),
in particular, all the expressions
$$\{ U^{\nu} (x) , U^{\mu} (y) \}^{(i)} \,\,\, =   $$
$$ = \,\,\, g^{\nu\mu i} \left( {\bf U} (x) \right) \,\,
\delta^{\prime} (x - y) \,\, + \,\, 
b^{\nu\mu i}_{\lambda} \left( {\bf U} (x) \right) 
U^{\lambda}_{x} \,\, \delta (x - y) $$
define in this case one-dimensional Poisson brackets 
compatible with each other.

 By analogy with the one-dimensional case, one can define the 
canonical form of a non-degenerate Poisson bracket (\ref{GenBr}) 
as a constant bracket
$$\{ n^{\nu} ({\bf x}) , n^{\mu} ({\bf y}) \} 
\,\,\, =  \,\,\, \eta^{\nu\mu i}  \,\,
\delta_{x^{i}} ({\bf x} - {\bf y}) \,\,\, ,$$
where all $\, \eta^{\nu\mu i} \, = \, {\rm const} \, $.
 
 It is easy to see that all the functionals
$$N^{\nu} \,\,\, = \,\,\, \int  n^{\nu} ({\bf x}) \, d^{n} x $$
reprsent in this case annihilators of the bracket (\ref{GenBr}).
 
 In contrast to the one-dimensional case, however, the 
nondegeneracy of the bracket (\ref{GenBr}) here in the general 
case turns out to be insufficient for the possibility of its 
reduction to the canonical form by means of some point change 
of coordinates, and some additional general conditions are 
required. In particular (see \cite{MokhovMnogomernSkob2}), 
a non-degenerate bracket (\ref{GenBr}) can be rediced to a 
constant form in some coordinates
$\, {\bf n} = {\bf n} ({\bf U}) \, $, 
if at least one of the metrics 
$\, g^{\nu\mu i_{0}} ( {\bf U} ) \, $ 
forms non-singular pairs with all other metrics, that is, 
the roots of any of the equations
$${\rm det} \,\, \Big( g^{\nu\mu i_{0}} ( {\bf U} )
\,\, - \,\, \lambda \,  g^{\nu\mu i} ( {\bf U} )
\Big) \,\,\, = \,\,\, 0 \,\,\, , \quad i \neq i_{0} $$
are different from each other.

 The above condition is generic, however, it is often 
violated in important examples. In particular, the Poisson 
brackets, corresponding to the Lie algebras of vector fields in
$\, \mathbb{R}^{n} \, $ ($N = n$, $n \geq 2$): 
$$\{p^{i} ({\bf x}) , p^{j} ({\bf y}) \} \,\,\, =  \,\,\, 
p^{i} ({\bf x}) \, \delta_{x^{j}} ({\bf x} - {\bf y}) 
\,\, - \,\, p^{j} ({\bf y}) \, 
\delta_{y^{i}} ({\bf y} - {\bf x}) $$
and describing $n$ - dimensional hydrodynamics, cannot be 
reduced to a constant form. The same is true, in fact, for 
a number of other important examples of hydrodynamic brackets.

 In the most general case, we can assert, however (see
\cite{DubrNovMnogomernSkob,MokhovMnogomernSkob1,MokhovMnogomernSkob2}),
that any non-degenerate bracket (\ref{GenBr}) can be reduced 
to a linear (non-uniform) form
\begin{multline}
\label{LinMultiDimBr}
\{ U^{\nu} ({\bf x}) , U^{\mu} ({\bf y}) \} \,\,\, =   \\
 = \,\,\, \Big( \left( b^{\nu\mu i}_{\lambda} \, + 
b^{\mu\nu i}_{\lambda} \right) U^{\lambda} \, + \, g^{\nu\mu i}_{0}
\Big) \,\, \delta_{x^{i}} ({\bf x} - {\bf y}) \,\,\, +  \\
+ \,\,\,  
b^{\nu\mu i}_{\lambda} \, U^{\lambda}_{x^{i}} \,\, 
\delta ({\bf x} - {\bf y}) \,\,\, , 
\end{multline}
$$b^{\mu\nu i}_{\lambda} \, = \, {\rm const} \quad , \,\,\,
g^{\nu\mu i}_{0} \, = \, {\rm const} $$
using a point change of coordinates.

 Thus, it can be seen that the theory of non-degenerate 
Poisson brackets in the case of several space variables is 
connected in the most general case with the theory of 
infinite-dimensional Lie algebras. A number of important 
classification results related to the theory of multi-dimensional 
brackets (\ref{LinMultiDimBr}) and the corresponding Lie-algebraic 
structures were obtained in the work \cite{MokhovMnogomernSkob1}. 
On the whole, however, the complete problem of classification of 
such brackets has not yet been finally solved.

 It must be said that the diagonal form of the Poisson bracket 
in the case of many spatial variables no longer plays the 
important role that it plays in the one-dimensional case.
As follows from the results of \cite{MokhovMnogomernSkob2},
nevertheless, in the case of two spatial variables, both metrics
$\, g^{\nu\mu 1} ({\bf U}) \, $ and
$\, g^{\nu\mu 2} ({\bf U}) \, $ 
can be reduced to diagonal form by a transformation of coordinates, 
if they form a non-singular pair. In the case $\, n \geq 3 \, $, 
generally speaking, the simultaneous reduction of all metrics
$\, g^{\nu\mu i} ({\bf U}) \, $ to the diagonal form is impossible.

 As in the one-dimensional case, one can define the Liouville form
for multi-dimensional brackets (\ref{GenBr}), that has the form
\begin{multline*}
\{U^{\nu}({\bf x}), U^{\mu}({\bf y})\} \,\,\, =  \\
= \,\,\, \Big( \gamma^{\nu\mu i}({\bf U}) \, + \,
\gamma^{\mu\nu i}({\bf U}) \Big) \,\, 
\delta_{x^{i}}({\bf x} - {\bf y}) \,\,\, +  \\
+ \,\,\, 
\Big( \gamma^{\nu\mu i}({\bf U}) \Big)_{x^{i}} \,\, 
\delta ({\bf x} - {\bf y})
\end{multline*}
and corresponds to the situation when all the functionals
$$I^{\nu} \,\,\, = \,\,\, \int  U^{\nu} ({\bf x}) \, d^{n} x $$
commute with each other.

\vspace{1mm}

 In conclusion, we would like to consider another 
generalization of brackets of hydrodynamic type, namely, 
brackets containing phase variables
$$ \{ S^{\alpha} ({\bf x}) , S^{\beta} ({\bf y}) \} 
\,\,\, = \,\,\, 0  $$
\begin{equation}
\label{PhaseHydrBr}
\{ S^{\alpha} ({\bf x}) , U^{\nu} ({\bf y}) \} \,\,\, = \,\,\,
\omega^{\alpha\nu} ({\bf U}, {\bf S}_{\bf x}) \,\, 
\delta ({\bf x}-{\bf y})  
\end{equation}
\begin{multline*}
\{ U^{\nu} ({\bf x}) , U^{\mu} ({\bf y}) \} \,\,\, = \,\,\, 
g^{\nu\mu i} \left( {\bf U}, {\bf S}_{\bf x} \right) \,\,
\delta_{x^{i}} ({\bf x} - {\bf y}) \,\,\, +  \\
+ \,\, 
b^{\nu\mu i}_{\lambda} \left( {\bf U}, {\bf S}_{\bf x} \right) 
U^{\lambda}_{x^{i}} \,\, \delta ({\bf x} - {\bf y}) \,\, + \,\,
f^{\nu\mu ij}_{\alpha}  \left( {\bf U}, {\bf S}_{\bf x} \right)
S^{\alpha}_{x^{i}x^{j}} \delta ({\bf x}-{\bf y}) 
\end{multline*}

  As already indicated above, brackets of this type arise, 
for example, when averaging local Hamiltonian structures in the 
multi-dimensional case. In fact, such brackets are also 
encountered in many other cases, where the variables 
$\, {\bf S} ({\bf x}) \, $ can have very different meanings 
(classical or quantum phases, phases of the order parameter, etc.).
In particular, brackets of this type are repeatedly encountered 
in the work \cite{DzyaloshinskiiVolovik}. Even more general 
brackets of this type can also include brackets where the phase 
variables do not commute with each other, but correspond to a certain 
Lie algebraic structure (see \cite{DzyaloshinskiiVolovik}). 
As noted above, brackets of the described type can also be found 
in ``pure'' hydrodynamics, for example, when describing potential 
flows. Quite often, the densities of conservation laws are chosen 
as the variables $\, {\bf U} ({\bf x}) \, $, and the hydrodynamic 
part of the bracket (\ref{PhaseHydrBr}) has a Liouville form.

 We will be interested here in the general structure of brackets 
(\ref{PhaseHydrBr}), and in particular, in the possibility of 
reducing such brackets to some canonical forms close to those 
considered above. From the physical point of view, the phase 
variables $\, {\bf S} ({\bf x}) \, $ are distinguished, so it is 
natural to actually consider changes of coordinates that preserve 
the variables $\, {\bf S} ({\bf x}) \, $ and transforming only the 
remaining coordinates $\, {\bf U} ({\bf x}) \, $. It must be said,
however, that the values $\, S^{\alpha}_{\bf x} \, $
(``superfluid velocities'') are already variables of the 
hydrodynamic type and can naturally be used in transformations
of the ``hydrodynamic variables''
$$U^{\nu} \,\,\, \rightarrow \,\,\, {\tilde U}^{\nu} 
\left( {\bf S}_{\bf x}, {\bf U} \right) $$

 Here we consider in a sense a non-degenerate case when the 
frequency matrix $\, \omega^{\alpha\nu} \, $ has the full rank
$${\rm rank} \, || \, \omega^{\alpha\nu} 
\left( {\bf S}_{\bf x}, {\bf U} \right) || 
\,\,\, = \,\,\, m $$
($ N-m \, \geq \, m $). As can be shown (\cite{AvBr}), 
in this case it is always possible to switch to new 
hydrodynamic variables
$${\bf U} \,\,\, \rightarrow \,\,\, 
\big( {\bf Q}  \left( {\bf S}_{\bf x}, {\bf U} \right) ,
{\bf N} \left( {\bf S}_{\bf x}, {\bf U} \right) \big) $$
$$Q_{\alpha} ({\bf x}) \,\, = \,\, 
Q_{\alpha} \left( {\bf S}_{x^{1}}, \dots, {\bf S}_{x^{d}},
\, {\bf U} ({\bf x}) \right) \,\,\, , 
\quad \quad \alpha = 1, \dots , m $$
$$N^{l} ({\bf x}) \,\, = \,\,
N^{l} \left( {\bf S}_{x^{1}}, \dots, {\bf S}_{x^{d}},
\, {\bf U} ({\bf x}) \right) \, , 
\quad l = 1, \dots , N - 2 m \, , $$
in which the bracket (\ref{PhaseHydrBr}) becomes
$$ \{ S^{\alpha} ({\bf x}) , S^{\beta} ({\bf y}) \} 
\,\,\, = \,\,\, 0  \,\,\,\,\, , $$
$$\left\{ S^{\alpha} ({\bf x}) \, , \, 
Q_{\beta} ({\bf y}) \right\}
\,\, = \,\, \delta^{\alpha}_{\beta} \,\,
\delta ({\bf x} - {\bf y}) \,\,\,\,\, , $$
\begin{equation}
\label{AlmostForm}
\left\{ S^{\alpha} ({\bf x}) \, , \, 
N^{l}  ({\bf y}) \right\} \,\, = \,\, 0 \,\,\, , 
\end{equation}
$$\left\{ Q_{\alpha} ({\bf x}) \, , \, 
Q_{\beta} ({\bf y}) \right\}
\,\, = \,\, J_{\alpha\beta} \, [{\bf S}_{\bf x}, {\bf N}] \, 
({\bf x}, \, {\bf y}) \,\,\, , $$
$$\left\{ Q_{\alpha} ({\bf x}) \, , \, 
N^{l}  ({\bf y}) \right\}
\,\, = \,\, J_{\alpha}^{l} \, [{\bf S}_{\bf x}, {\bf N}] \,
({\bf x}, \, {\bf y}) \,\,\, , $$
$$\left\{  N^{l}  ({\bf x}) \, , \, 
N^{q}  ({\bf y}) \right\}
\,\, = \,\, J^{lq} \, [{\bf S}_{\bf x}, {\bf N}] \,
({\bf x}, \, {\bf y}) $$

 The commutators
$\, J_{\alpha\beta} \, [{\bf S}_{\bf x}, {\bf N}] \, 
({\bf x}, \, {\bf y}) $,
$\, J_{\alpha}^{l} \, [{\bf S}_{\bf x}, {\bf N}] \,
({\bf x}, \, {\bf y}) $ and 
$\, J^{lq} \, [{\bf S}_{\bf x}, {\bf N}] \,
({\bf x}, \, {\bf y}) \, $ 
are given by expressions of hydrodynamic type similar to 
those presented in (\ref{PhaseHydrBr}), here, however, they 
depend only on the variables $\, {\bf S} ({\bf x}) \, $ and 
$\, {\bf N} ({\bf x}) \, $. It is also not diffucult to show
that the functionals
$\, J^{lq} \, [{\bf S}_{\bf x}, {\bf N}] \,
({\bf x}, \, {\bf y}) \, $ define a Poisson bracket on the 
space of fields $\, {\bf N} ({\bf x}) \, $ for any fixed 
values of $\, {\bf S} ({\bf x}) \, $.

 As can be easily seen, the variables
$\, Q_{\alpha} ({\bf x}) \, $ and $\, N^{l} ({\bf x}) \, $
are defined up to the transformations
$$Q_{\alpha} ({\bf x}) \,\,\,\,\, \rightarrow \,\,\,\,\,
Q_{\alpha} ({\bf x}) \,\, + \,\, f_{\alpha}
\left( {\bf S}_{\bf x}, \, {\bf N} ({\bf x}) \right) 
\,\,\,\,\, , $$ 
$$N^{l} ({\bf x}) \,\,\,\,\, \rightarrow \,\,\,\,\,
N^{\prime l} \left( {\bf S}_{\bf x}, \,
{\bf N} ({\bf x}) \right) $$
where
$${\rm det} \,\, \left| \left|
{\partial N^{\prime l} \over \partial N^{k}}
\right| \right| \,\,\, \neq \,\,\, 0 $$
 
 Quite often, in fact, we are interested in the situation 
when the variables $\, N^{l} ({\bf x}) \, $ do not appear
($N = 2m$), and the bracket (\ref{PhaseHydrBr})
is reduced to the form
$$ \{ S^{\alpha} ({\bf x}) , S^{\beta} ({\bf y}) \} 
\,\,\, = \,\,\, 0  \,\,\,\,\, , $$
\begin{equation}
\label{VarAdmBr}
\left\{ S^{\alpha} ({\bf x}) \, , \, 
Q_{\beta} ({\bf y}) \right\}
\,\, = \,\, \delta^{\alpha}_{\beta} \,\,
\delta ({\bf x} - {\bf y}) \,\,\,\,\, , 
\end{equation}
\begin{multline*}
\left\{ Q_{\alpha} ({\bf x}) \, , \,
Q_{\beta} ({\bf y}) \right\}
\,\,\,\,\, =  \,\,\,\,\, \Omega_{\alpha\beta}^{i} 
( {\bf S}_{\bf x}) \,\,\, 
\delta_{x^{i}} ({\bf x} - {\bf y}) \,\,\, +  \\
+ \,\,\, 
\Gamma_{\alpha\beta\gamma}^{ij} ( {\bf S}_{\bf x} ) \,\,
S^{\gamma}_{x^{i}x^{j}} \,\,\, \delta ({\bf x} - {\bf y}) 
\end{multline*}
$(\Gamma_{\alpha\beta\gamma}^{ij} \, \equiv \, 
\Gamma_{\alpha\beta\gamma}^{ji})$.

 The Jacobi identities 
$$\left\{ \left\{ Q_{\alpha} ({\bf x}) \, , \,
Q_{\beta} ({\bf y}) \right\} \, , \,   
Q_{\gamma} ({\bf z}) \right\} \,\,\,\,\, + \,\,\,\,\,
c.p. \,\,\,\,\, \equiv \,\,\,\,\, 0 $$
now give the relations
$${ \delta J_{\alpha\beta} [{\bf S}_{\bf x}] ({\bf x}, {\bf y})
\over \delta S^{\gamma} ({\bf z})}  + 
{ \delta J_{\beta\gamma} [{\bf S}_{\bf x}] ({\bf y}, {\bf z})
\over \delta S^{\alpha} ({\bf x})}  + 
{ \delta J_{\gamma\alpha} [{\bf S}_{\bf x}] ({\bf z}, {\bf x})   
\over \delta S^{\beta} ({\bf y})} \,\, \equiv \,\, 0 $$
for the functionals
$\, J_{\alpha\beta} [{\bf S}_{\bf x}] ({\bf x}, {\bf y})$,
which mean that the 2-form
$$\int \, J_{\alpha\beta} [{\bf S}_{\bf x}] \, ({\bf x}, {\bf y}) \,\,\,
\delta S^{\alpha} ({\bf x})  \wedge  
\delta S^{\beta} ({\bf y}) \,\,\, d^{n} x \, d^{n} y $$
is closed on the space of fields
$\, (S^{1} ({\bf x}), \dots, S^{m} ({\bf x}))$.

 The brackets (\ref{VarAdmBr}) belong to the general class of 
brackets, named in \cite{HamFormMultAnMorseTheory} as 
``variationally admissible''. The variationally admissible 
form of Poisson brackets is directly related to a possibility 
of a Lagrangian description of the corresponding dynamical 
systems and, as was shown in \cite{HamFormMultAnMorseTheory}, such 
brackets generally lead to a nontrivial Lagrangian representation 
of Hamiltonian systems, where the Lagrange functional is 
in fact a 1-form, which has nontrivial topological properties.

 In our case, we need to remember, in fact, that when reducing 
the bracket (\ref{VarAdmBr}) to the canonical form, we are limited 
only to the transformations of the ``hydrodynamic type'' presented 
above. As it was shown in \cite{AvBr}, any bracket (\ref{VarAdmBr}) 
can be locally reduced to the canonical form
$$ \{ S^{\alpha} ({\bf x}) , S^{\beta} ({\bf y}) \} 
\,\,\, = \,\,\, 0  \,\,\,\,\, , $$
$$\left\{ S^{\alpha} ({\bf x}) \, , \, 
Q_{\beta} ({\bf y}) \right\}
\,\, = \,\, \delta^{\alpha}_{\beta} \,\,
\delta ({\bf x} - {\bf y}) \,\,\,\,\, , $$
$$\left\{ Q_{\alpha} ({\bf x}) \, , \,
Q_{\beta} ({\bf y}) \right\}
\,\,\,\,\, =  \,\,\,\,\, 0 $$
using the transformation
$$Q_{\alpha} ({\bf x}) \,\,\,\,\, \rightarrow \,\,\,\,\,
Q_{\alpha} ({\bf x}) \,\, + \,\, f_{\alpha}
\left( {\bf S}_{\bf x} \right) $$ 
 
 As a consequence, in this case the corresponding Hamiltonian 
system can also be written in Lagrangian form with the Lagrangian 
of the ``hydrodynamic type''
\begin{multline*}
\delta  \int  \Big[ \, Q_{\alpha} ({\bf X}) \, S^{\alpha}_{t}
\,\,\, - \,\,\, \langle P_{H} \rangle \left( {\bf S}_{\bf x}, 
\, {\bf Q} ({\bf X}) \right) \Big] \,\,
d^{n} x \, d t  \,\,\,  =  \,\,\,  0  
\end{multline*}

 As for the general brackets (\ref{AlmostForm}), here we can also 
raise the question of further reducing them to their canonical form 
and, in particular, of separating brackets for the variables
$\, \left( {\bf S} ({\bf x}), {\bf U} ({\bf x}) \right) \, $ and
$\, {\bf N} ({\bf x}) \, $. In fact, such a possibility often arises 
in special examples and, in particular, in the theory of slow 
modulations for multi-dimensional systems. It can be shown, however, 
that in the most general case it is impossible to reduce the brackets 
(\ref{AlmostForm}) to such a canonical form using a transformation of 
the ``hydrodynamic type'' (\cite{AvBr}).

\vspace{1mm}

 In conclusion, we note also that the brackets (\ref{PhaseHydrBr}) 
often have a natural continuation to the extended phase space, 
in which the variables $\, v^{\alpha}_{i} = S^{\alpha}_{x^{i}} \, $  
can be considered completely independent. Extensions of this type 
can to some extent be naturally called the ``vortizations'' of
the brackets (\ref{PhaseHydrBr}). Such continuations naturally 
arise not only, for example, in hydrodynamics, during the transition 
from potential to vortex flows, but also, for example, in the 
description of the motion of superfluids carrying quantum vortex 
structures inside their volume, etc. Important examples of such
``vortizations'' of the brackets like (\ref{PhaseHydrBr}) are given 
in the work \cite{DzyaloshinskiiVolovik}.

\section{Conclusion}
\setcounter{equation}{0}

 This paper provides a brief overview of the Poisson brackets 
of hydrodynamic type and their special generalizations. We consider 
questions related to various forms of such brackets and, 
in particular, to representations generalizing the canonical forms 
of Poisson brackets in the situation under consideration.
The connection between brackets of hydrodynamic type and the 
theory of Lie algebras in the cases of one and several spatial 
variables is considered. The connection of the considered 
structures with the theory of integration of systems of 
hydrodynamic type in the one-dimensional case is described 
in particular detail.

\end{document}